\documentclass[a4paper,fleqn,usenatbib]{mnras}
\usepackage[T1]{fontenc}
\usepackage{ae,aecompl}

\usepackage{graphicx}
\usepackage{amsmath}
\usepackage{amssymb}
\usepackage{bm}

\usepackage{aas_macros}

\newcommand{\vecb}[1]{{\bm{\mathrm{#1}}}}
\newcommand{\Dfrac}[2]{\frac{d#1}{d#2}}

\title[Recovery of the spectrum of turbulence]{On the technique for the recovery of the spectrum of turbulence in astrophysical disks}

\author[D. V. Bisikalo et al.]{
D. V. Bisikalo,$^{1}$\thanks{E-mail: bisikalo@inasan.ru}
E. P. Kurbatov,$^{1}$
Ya. N. Pavlyuchenkov,$^{1}$
A. G. Zhilkin$^{1,2}$
\newauthor{and P. V. Kaygorodov$^{1}$}
\\
$^{1}$Institute of Astronomy of the RAS, Moscow, Russia\\
$^{2}$Chelyabinsk State University, Chelyabinsk, Russia
}

\date{Accepted XXX. Received YYY; in original form ZZZ}

\pubyear{2016}

\begin{document}
\label{firstpage}
\pagerange{\pageref{firstpage}--\pageref{lastpage}}
\maketitle

\begin{abstract}
We present a method that can be used to recover the spectrum of turbulence from observations of optically thin emission lines formed in astrophysical disks. Within this method we analyze how line intensity fluctuations depend on the angular resolution of the instrument, used for the observations. The method allows us to restore the slope of the power spectrum of velocity turbulent pulsations and estimate the upper boundary of the turbulence scale.
\end{abstract}




\section{Introduction}

Accretion disks~-- mainly protoplanetary disks and those of binary stars~-- are of extreme interest, since they demonstrate the great variety of observational manifestations and allow studying a big number of physical processes. The main question about the mechanism of the angular momentum transfer in the disks has not been completely solved yet, though the disks have been attracting the interest of many scientists for several decades. From observations one can derive the rate of accretion onto the central object, which enables determining the rate of angular momentum loss in the disk. If we suppose that the angular momentum transfer is driven solely by the material viscosity in the disk, then we find that the viscosity coefficient in the disks should be anomalously high, by $9 \-- 11$ orders of magnitude higher than the coefficient of molecular viscosity.

There have been a number of physical processes proposed that could drive the angular momentum transfer: tidal interaction \citep{Papaloizou1977MNRAS.181..441P,Lin1990ApJ...365..748L, Dgani1994ApJ...436..270D}; spiral shock waves \citep{Michel1984ApJ...279..807M,Sawada1986MNRAS.219...75S,Spruit1987A&A...184..173S,Syer1993MNRAS.262..749S}; convection \citep{Bisnovatyi-Kogan1976PAZh....2..489B,Paczynski1976ComAp...6...95P,Shakura1978A&A....62..179S,Lin1980MNRAS.191...37L}; accretion disk wind and jets \citep{Blandford1982MNRAS.199..883B,Lubow1994MNRAS.268.1010L,Cao2002A&A...385..289C, Lynden-Bell2003MNRAS.341.1360L}; angular momentum transfer by propagating waves \citep{Lin1979MNRAS.186..799L,Lubow1981ApJ...245..274L,Vishniac1989ApJ...347..435V}; and various instabilities that can result in the most probable mechanism, well-developed turbulence. The comparison of probable mechanisms show that the high accretion rate in the disks may be explained only by the turbulent viscosity \citep{Shakura1972AZh....49..921S,Shakura1973A&A....24..337S,Lynden-Bell1974MNRAS.168..603L}
(see also a review by \citet{Shakura2014PhyU..184..445S}).

To answer the question, whether the angular momentum transfer is driven by developed turbulence in the disk, and, if so, to understand the nature of the turbulent viscosity, one should be able, if possible, to derive the parameters of turbulence from observational data. For the protoplanetary disks, it has been able to determine mostly the rough parameters of the turbulence rather than the refined energy spectrum obtained for interstellar medium. In part, it is done by considering the variations of accretion rate and/or estimating specific lifetimes and sizes of protoplanetary disks. In terms of $\alpha$-parameter, proposed by \citet{Shakura1973A&A....24..337S}, the estimated value of the turbulent viscosity varies in a wide range $\alpha = 10^{-3} \dots 10^{-1}$, see e.g. \cite{Hartmann1998ApJ...495..385H}. Another possibility to study turbulence in protoplanetary disk is to use observations of molecular lines. A general method to derive the turbulent velocity, averaged over the disk, by fitting the multi-parametric model to interferometric spectra maps was presented in \cite{Guilloteau1998A&A...339..467G}. With the help of this model the authors of \citet{Dartois2003A&A...399..773D} and \citet{Guilloteau2012A&A...548A..70G} estimated the turbulent line width using the observations of different molecular transitions in the DM~Tau protoplanetary disk. Their results show that the turbulence in this disk is sub-sonic with specific velocities of $0.1 \dots 0.5$ of Mach number.

Turbulent motions in gas can have different values of kinetic energy on different spatial scales. Due to the Doppler effect, the turbulent velocity pulsations result in the fluctuations of radiation intensity in the frequency space as well as in the volume of the turbulent medium. In particular, within the Kolmogorov approach the large-scale vortices carry the major fraction of the energy of turbulent pulsations, while the energy of small-scale vortices is less but they are more numerous. Thus, the different parts of the turbulent energy spectrum causes the different responses in the intensity fluctuations. This potentially enables us to restore the power spectrum of the velocity pulsations in the turbulent medium using analysis of the spectral and/or spatial distribution of the radiation intensity.

A possible method for extracting the power index of the turbulence was proposed by \citet{Lazarian2006ApJ...652.1348L} (see also \citet{Lazarian2004ApJ...616..943L} and \citet{Lazarian2000ApJ...537..720L}). The method utilizes fluctuations of the intensity on different frequencies in a single emission line and in the different lines of sight. The authors showed a connection between the structure function of the spectral intensity distribution and the statistical properties of the turbulent velocity as well as the density pulsations. Using the structure function and assuming Gaussian distribution for turbulent velocity it becomes possible to extract the power spectrum of the turbulence.

In the present paper we present an alternative approach to the problem of the extraction of turbulent properties from the spectral observations. As a basic quantity to express the characteristics of the turbulence, we choose the variance of the fluctuations of the center of the observed emission line. We show that this measure strongly depends on angular resolution of the telescope. The telescope's aperture acts as a filter for angular modes of the image, smoothing out the small-scale angular fluctuations.  This filtering reduces the spatial scale of turbulent velocity pulsations which are reflected in the emergent intensity fluctuations. Hence, it is possible to formulate a relation between the angular resolution of the telescope and the amount of the turbulent energy at a corresponding spatial scale.

The paper's structure is as follows. In Section 2 we briefly describe the model, explaining how turbulent fluctuations occur in an emission line, discuss the influence of the telescope's resolution on the observed properties of the fluctuations, and propose a method allowing us to recover the parameters of the turbulent medium by analyzing the statistical properties of the intensity fluctuations. In Section 3 we perform numerical analysis of the proposed method. Section 4 is focused on various features of the method and we discuss a number of issues, concerned with the interpretation of the parameters of the turbulent medium, in particular, a possible difference between the observed spectrum of turbulence and `classic' Kolmogorov spectrum. In the Conclusions we summarize the results of our study. Appendixes contain the detailed mathematics formalization of the model.

\section{Recovery of parameters of turbulent medium from intensity fluctuations}
\label{sec:reconstruction}

In this section we analyze an optically thin emission line that forms in a turbulent layer of finite height along vertical direction. We assume that physical parameters are the same in horizontal direction except temperature which can vary. We discuss the application and modification of this model to real accretion disks in Discussion.
One can state that the profile of an optically thin emission line is a sum of profiles, forming in elementary emission events and shifted due to the Doppler effect. The reason of the Doppler shift is the thermal and turbulent motions in the medium. The specific time, required by a gas-dust disk to reach thermal equilibrium, is usually smaller than it's dynamic timescale \citep{Vorobyov2014ARep...58..522V}. Therefore, we can neglect temporal fluctuations of thermal line broadening which otherwise could appear due to dynamical and radiative heating/cooling.

In real gas-dust disks there are radial and vertical temperature and density gradients. These gradients may affect properties of the turbulence. However, for the sake of simplicity, hereafter we assume the turbulence to be uniform and isotropic in the medium. As the magnitude of the velocity pulsations may vary over different scales and different lines of sight, the turbulent contribution to the line profile should be treated as a random. Thus, the optically thin emission line in a macroscopic volume of the turbulent medium is the sum of thermal profiles that are randomly Doppler-shifted due to the turbulent motions of the gas. Further we will assume the turbulence is subsonic, which will allow us to neglect pulsations of the gas density.

It is known that large vortices usually contain the main fraction of turbulent energy. These vortices produce high velocity shifts that cause intensity fluctuations in the line wings. Velocity pulsations in the line core are driven by small vortices. On the other hand, the number of small vortices is much bigger than that of the large vortices. Thus, the effect of the line broadening by small vortices is averaged. This results in the smooth line core, but the line wings nonetheless has strong fluctuations. Hence, the structure of the intensity fluctuations in a line profile gives us information about the energy distribution over various spatial scales in the turbulent medium.

Any telescope, including a radio-interferometer, acts as a filter of spatial harmonics that effectively averages the image over small-scale components in accordance with its beam pattern. Since the turbulence in astrophysical objects is, as a rule, three-dimensional the spatial filtering in the image plane corresponds to the spatial filtering along the line of sight. This means that, by conducting spectral observations at different spatial resolutions, one obtains fluctuation samples on different velocity scales. The less angular resolution of the telescope (wide beam pattern), the smaller the number of small-scale velocity pulsations in the spectral map. Thus, when having a sufficiently comprehensive sample of emission lines observed at different angular resolutions, one can potentially study the statistical properties of velocity pulsations on various spatial scales.

\subsection{Formation of turbulent intensity fluctuations in an optically thin emission line}

We mentioned above that the shape of an optically thin line profile is governed by both the thermal and turbulent motions in gas as well as the number density distribution in the medium. We use the following expression for the intensity. Let $I(v, \vecb{R})$ be the intensity at the radial velocity $v$, for the line of sight $\vecb{R} = (x, y)$ along vertical direction, where $(x, y)$ are the coordinates in the image plane.
\begin{equation}
  \label{eq:intensity_exact}
  I(v, \vecb{R})
  = \varepsilon \int dz\,n(z)\,\phi[ v - u(\vecb{r}), \sigma_\mathrm{th}(\vecb{r}) ]  \;,
\end{equation}
where $\varepsilon$ is the emissivity of a single particle; $n$ is the number density of particles, we allow it to vary in a $z$-direction only; $\phi[v-u, \sigma_\mathrm{th}]$ is the Gaussian profile centered at $v = u$ and having the standard deviation equal to the thermal broadening $\sigma_\mathrm{th}$; $u(\vecb{r})$ is the radial velocity at a point with coordinates $\vecb{r} = (x, y, z)$. The integration performed along the line of sight.

If one assumes the turbulence to be three-dimensional and isotropic, the radial velocity distribution $u(\vecb{r})$ should completely describe the field of turbulent pulsations. Recovering of the radial velocity field from observed line profiles is, strictly speaking, an ill-posed inverse problem. However, certain properties of turbulent pulsations are encoded in the moments of the spectral intensity distribution.

We consider the column density is constant hence the total intensity does not depend on the direction:
\begin{equation}
  \int dv\,I(v, \vecb{R}) = \varepsilon N \equiv \text{const}  \;,
\end{equation}
where the column number density $N$ is
\begin{equation}
  N(\vecb{R}) = \int dz\,n(z)  \;.
\end{equation}
The first moment of the intensity distribution may not be equal to zero as it depends on the turbulent velocity pulsations averaged along the line of sight:
\begin{equation}
  \label{eq:momentum1}
  \overline{v}(\vecb{R})
  = \frac{1}{\varepsilon N} \int dv\,v\,I(v, \vecb{R})
  = \int dz\,\frac{n(z)}{N}\,u(\vecb{r})  \;.
\end{equation}
The Doppler shift, caused by turbulent pulsations, adds to the line broadening along with the thermal motions of the particles. It is easy to show that the second moment of a line profile is determined by the thermal broadening and the specific distribution of radial velocities:
\begin{align}
  \label{eq:momentum2}
  \overline{v^2}(\vecb{R})
  &{}= \frac{1}{\varepsilon N} \int dv\,v^2 I(v, \vecb{R})  \\
  \label{eq:momentum2_b}
  &{}= \int dz\,\frac{n(z)}{N}\,[ \sigma_\mathrm{th}^2(\vecb{r}) + u^2(\vecb{r}) ]  \;.
\end{align}
The centered second moment (calculated with respect to the line center \ref{eq:momentum1}) is:
\begin{align}
  \label{eq:momentum2_centered}
  \overline{[v - \overline{v}(\vecb{R})]^2}
  &{}= \frac{1}{\varepsilon N} \int dv\,[v - \overline{v}(\vecb{R})]^2 I(v, \vecb{R})  \\
  &{}= \overline{v^2}(\vecb{R}) - [\overline{v}(\vecb{R})]^2  \;.
\end{align}

Since the field of radial velocities is random, the first moment of a line profile is also a random quantity. Thus, the shift of the emission line center (\ref{eq:momentum1}) may have two components: a systematic Doppler shift (caused by the motion of the source as a whole) and a random shift, whose absolute value and sign are determined by the field of turbulent velocity pulsations. The same effect influences the line width given by (\ref{eq:momentum2}) and (\ref{eq:momentum2_centered}).

By analyzing a wide sample of observed line profiles one can find a dependence between the statistical properties of line intensity fluctuations and the statistical properties of turbulent velocity pulsations. Let $\langle \cdot \rangle$ denote the operation of averaging over the ensemble of velocity pulsations. For the sake of convenience we assume that the mean value of the pulsations is zero. We also neglect pulsations of the number density assuming the turbulence is sub-sonic. This results in the zero value of the averaged radial velocity of line profile:
\begin{equation}
  \langle \overline{v}(\vecb{R}) \rangle
  = \int dz\,\frac{n(z)}{N}\,\langle u(\vecb{r}) \rangle \equiv 0  \;.
\end{equation}
Let us find the variance of the turbulent velocity as the ensemble average given that the three-dimensional turbulence is uniform and isotropic:
\begin{equation}
  \sigma_\mathrm{turb}^2 = \langle u^2(\vecb{r}) \rangle \equiv \text{const}  \;.
\end{equation}
The last assumption implies that the average contribution of turbulence to the line broadening is solely determined by the variance of the turbulent velocity and does not depend on the line-of-sight direction. But the thermal contribution may vary from point to point, hence the total line broadening becomes
\begin{equation}
  \label{eq:mean_momentum2}
  \langle \overline{v^2}(\vecb{R}) \rangle
  = \int dz\,\frac{n(z)}{N}\,\sigma_\mathrm{th}^2(\vecb{r}) + \sigma_\mathrm{turb}^2  \;.
\end{equation}
The ensemble average of the centered second moment (\ref{eq:momentum2_centered}) is a difference between (\ref{eq:mean_momentum2}) and the variance of the line center fluctuations:
\begin{equation}
  \langle \overline{[v - \overline{v}(\vecb{R})]^2} \rangle
  = \langle \overline{v^2}(\vecb{R}) \rangle
    - \langle [\overline{v}(\vecb{R})]^2 \rangle  \;,
\end{equation}
where the second term is determined by the spatial correlation function of the radial velocity:
\begin{equation}
  \langle [\overline{v}(\vecb{R})]^2 \rangle
  = \int dz\,\frac{n(z)}{N} \int dz'\,\frac{n(z')}{N}\,
      \langle u(\vecb{r})\,u(\vecb{r}') \rangle  \;.
\end{equation}
It is seen that this quantity does not depend on the spatial temperature distribution but the density distribution and the spatial statistics of the turbulence. It appears that this value, when measured by the filtered intensity distribution, is sensitive to the filter's scale. From the observational point of view such the filtration is equivalent to the action of aperture of the telescope. Below we will show how properties of the filtered line center fluctuations are connected to the properties of the turbulent medium.

\subsection{The recovery of properties of turbulent medium from intensity fluctuations}

Modern instruments, just put in operation, (e.g. ALMA) enable us to conduct high-resolution radio-interferometric observations of optically thin molecular transitions. By reducing the resolution, we can filter out certain spatial scales and, using the proposed technique, derive slices of the turbulent characteristics on various spatial scales.

We can model the variations of the beam pattern by filtering the intensity field in the image plane (see Appendix B). Let us use the term $I_L(v, \vecb{R})$ to denote an intensity distribution, observed by a telescope with a beam pattern that corresponds to the filter $B_L(\vecb{R})$ with the specific width $L$:
\begin{equation}
  \label{eq:intensity_filtered}
  I_L(v, \vecb{R})
  = \int d^2R'\,B_L(\vecb{R}' - \vecb{R})\,I(v, \vecb{R}')  \;,
\end{equation}
where the filter satisfies the normalization $\int d^2R\,B_L(\vecb{R}) = 1$. By assuming the statistical uniformity, we can show that the ensemble average of the line profile depends neither on the filter nor on the line-of-sight direction (strictly speaking this statement is correct in the case of sub-sonic turbulence only, when one can neglect density fluctuations (see Discussion)),
\begin{equation}
  \langle I_L(v, \vecb{R}) \rangle = \langle I(v, \vecb{R}) \rangle  \;.
\end{equation}

Let us find the expression for the center of a line (\ref{eq:momentum1}), to which we applied the spatial filtering procedure (\ref{eq:intensity_filtered}) with a filter's width $L$:
\begin{align}
  \label{eq:center_velocity}
  \overline{v}_L(\vecb{R})
  &{}= \frac{1}{Q N} \int dv\,v\,I_L(v, \vecb{R})  \\
  &{}= \int d^2R'\,B_L(\vecb{R}'-\vecb{R})\,\overline{v}(\vecb{R})  \;.
\end{align}
Note that this quantity, as well as (\ref{eq:momentum1}), does not depend on the temperature distribution in the medium, since it disappeared when integrating over the velocity in (\ref{eq:center_velocity}).

Denote the variance of the line's center as $U_L$. It is easy to show that it does not depend on the line-of-sight direction, since turbulence is spatially uniform:
\begin{multline}
  \label{eq:variance_momentum1}
  U_L = \left\langle \overline{v}_L^2 \right\rangle
  = \int d^2R\,B_L(\vecb{R}) \int d^2R'\,B_L(\vecb{R}')  \\
  \times \int dz\,\frac{n(z)}{N} \int dz'\,\frac{n(z')}{N}
    \,\langle u(\vecb{r})\,u(\vecb{r}') \rangle  \;.
\end{multline}
However, it is essential that $U_L$ depends on the filter width $L$. It is determined by the spatial correlation function of velocity pulsations $\langle u(\vecb{r})\,u(\vecb{r}') \rangle$ that may be expressed via the power spectrum of turbulent pulsations $P(k)$:
\begin{equation}
  \label{eq:velocity_correlation}
  \langle u(\vecb{r})\,u(\vecb{r}') \rangle
  = \int d^3k\,e^{-i \vecb{k} (\vecb{r} - \vecb{r}')} P(k)  \;.
\end{equation}
We can assume that within the inertial interval the dependence of the power spectrum on the wave number is a power law: $P(k) \propto k^{-\alpha-2}$, where $\alpha > 1$. For instance, for the Kolmogorov turbulence $\alpha = 5/3$. The finite angular resolution filters out wave numbers in (\ref{eq:velocity_correlation}) that are greater than $L^{-1}$ (or spatial scales smaller than $L$). The specific energy of turbulent pulsations on the scale of wave numbers $k$ is approximately $k^3 P(k) \propto k^{1-\alpha}$, which means that pulsations on larger spatial scales carry a larger amount of energy. Thus, the main contribution to the line intensity fluctuations obviously comes from turbulent pulsations on the spatial scale corresponding to the angular resolution of the telescope.

An approximate expression for $U_L$ can be as follows (see Appendix B):
\begin{equation}
  \label{eq:variance_approx}
  U_L \approx
  \frac{\alpha - 1}{2\alpha}\,\frac{h_\mathrm{turb}}{h}\,\sigma_\mathrm{turb}^2
    \left[ 1
      - \left( \frac{L}{h_\mathrm{turb}} \right)^\alpha \right]  \;,
\end{equation}
where $h_\mathrm{turb}$ is the largest turbulent scale, and $h$ is the characteristic thickness of the turbulent layer (it depends on the number density distribution). This function monotonically decreases with the increasing ratio $L/h_\mathrm{turb}$. In particular, given the Kolmogorov slope $\alpha = 5/3$ and $h_\mathrm{turb} = h/2$, we have the maximum value of $U_L$ at $L = 0$ (infinitely high resolution) that can be estimated as $0.1 \sigma_\mathrm{turb}^2$. At extremely large $L$, the approach we use to derive (\ref{eq:variance_approx}) is incorrect, but one can expect $U_L$ to monotonically decrease with growing $L$ and become close to zero at $L \gtrsim h_\mathrm{turb}$.

We suppose that it is possible to estimate the power $\alpha$ by measuring $U_L$ as a function of telescope's resolution. This value can be derived from the logarithmic slope of the following function:
\begin{equation}
  \label{eq:variance_approx_pow}
  \max_L(U_L) - U_L \approx
  \frac{\alpha - 1}{2\alpha}\,\frac{h_\mathrm{turb}}{h}\,\sigma_\mathrm{turb}^2
    \left( \frac{L}{h_\mathrm{turb}} \right)^\alpha  \;.
\end{equation}
Thus, we propose the following method to restore the power spectrum:
\begin{enumerate}
  \item We calculate the variance $U_L$ of the line center's radial velocity for observations with various angular resolution $L$.
  \item We approximate the dependence of $U_L$ on $L$ by the expression \eqref{eq:variance_approx_pow} and derive $\alpha$.
\end{enumerate}

We have shown above that the variance of fluctuations of the line center averaged over a sufficiently comprehensive sample, depend on the resolution of the telescope, Eq. (\ref{eq:variance_momentum1}). We also may write out the ensemble-averaged line width:
\begin{multline}
  \label{eq:mean_momentum2_centered}
  \langle \overline{[v - \overline{v}(\vecb{R})]^2} \rangle
  = \int d^2R\,B_L(\vecb{R}) \int dz\,\frac{n(z)}{N}\,\sigma_\mathrm{th}^2(\vecb{r})  \\
    + \sigma_\mathrm{turb}^2 - U_L  \;.
\end{multline}
One can see that the variance of the position of the line center (\ref{eq:variance_momentum1}) decreases with the growing filter width $L$ (lower resolution), while the squared line width (\ref{eq:mean_momentum2_centered}) increases. Both these values can be used to estimate the power spectrum according to (\ref{eq:variance_approx_pow}), but to use expression (\ref{eq:mean_momentum2_centered}) we first should estimate the full line width (\ref{eq:mean_momentum2}) measured regardless of the fluctuations of its center. On the other hand the estimate of the full line width may be useful if one can, by some means, finds the value of thermal broadening $\sigma_\mathrm{th}$ distribution. In this case we can estimate the full turbulent line width $\sigma_\mathrm{turb}$ and, by using (\ref{eq:variance_approx}) with $L \to 0$, the largest scale of turbulence $h_\mathrm{turb}$.

The idea to extract properties of the turbulence from the intensity fluctuations is not new. \citet{Lazarian2006ApJ...652.1348L} proposed a technique for recovery the power spectrum of the velocity or density fluctuations using the spectrum of the intensity fluctuations measured along the velocity axis in the position-position-velocity data cubes. The intensity fluctuations spectrum itself is determined via the variance of difference of the intensity at different frequencies or, equivalently, velocities. This idea may be schematically represented as
\begin{equation}
  P(k_v)
  \propto \int dv\,\cos(k_v v) \int dv_+ \left\langle [I(v_1) - I(v_2)]^2 \right\rangle  \;,
\end{equation}
where $v_+ = (v_1 + v_2)/2$ and $v = v_1 - v_2$. The intensities related to the same line of sight, in case of high resolution limit, or averaged over a beam of a finite width. For optically thin emission line the intensity expressed similar to (\ref{eq:intensity_exact}) but the number density allowed to fluctuate arbitrarily. To perform the integration over $v_+$ the Gaussian distribution of the turbulent velocity pulsations was assumed. The authors showed that the power spectrum $P(k_v)$ may be expressed via the structure or correlation functions for the density and velocity fluctuations. Assuming e.g. the density structure function it is possible to extract the turbulent velocity power spectrum from the observations via $P(k_v)$. Great advantage of this method is that it allows for the density fluctuations and the finite absorption coefficient.

The method described in the present paper utilizes a completely different idea. We find that the variance of the first moment of the optically thin line profile depends on the angular resolution of the telescope. The variance can be expressed via the spatial correlation function of the velocity pulsations which is connected to the turbulent power spectrum. Thus, the filtering effect of the telescope's aperture makes possible to extract parts of the power spectrum of turbulence. As the advantage of this method can be noted that the first moment of the line profile determined by the turbulence, not the thermal distribution in the medium, hence it is free from uncertainties caused by the temperature gradients.

\section{Numerical tests}

To test the proposed method we have calculated two models. In the first model we represent the turbulence by a Gaussian random field of velocity fluctuations. In the second model we represent the turbulence by a set of vortices with rigid-body rotation. In both cases the medium is assumed isothermal. In the tests we calculated the power spectrum of turbulent velocity pulsations directly (from the velocity distribution) and using the proposed method.

\subsection{Gaussian field of velocity fluctuations}

\begin{figure*}
  \begin{center}
    \includegraphics{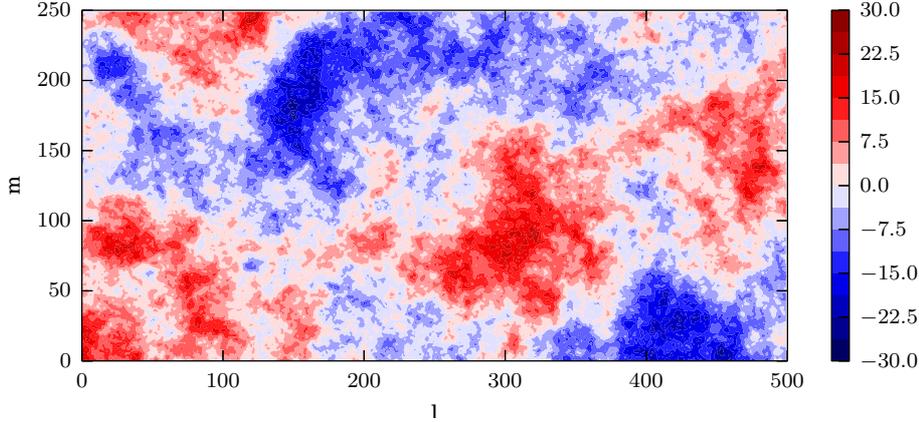}
  \end{center}
  \caption{The radial velocity field in the test model, where the turbulent medium is represented by a Gaussian field of velocity fluctuations with a Kolmogorov power spectrum ($\alpha = 1.67$).}
  \label{fig:test_gaussian_field_vel}
\end{figure*}

Let us define the turbulent field of radial velocities as a sum of independent realizations of Fourier harmonics on a 2-D discrete grid with the dimension of $N_x \times N_z$. We look at the system along the Z-axis. We derive the distribution of radial velocity over the grid by applying the inverse Fourier transform
\begin{equation}
  u_{lm} = \frac{1}{N_x N_z} \sum_{q = 0}^{N_x - 1} \sum_{p = 0}^{N_z - 1}
    \exp\!\left( i\,\frac{2\pi}{N_x}\,lq + i\,\frac{2\pi}{N_z}\,mp \right)\,\tilde{u}_{qp}
\end{equation}
for harmonics $\tilde{u}_{qp}$, where spatial indexes $l,m$ and wave indexes $q, p$ vary as
\begin{equation}
  l, q = \overline{0, (N_x - 1)}  \;,\qquad
  m, p = \overline{0, (N_z - 1)}  \;.
\end{equation}
Each pair of wave indexes corresponds to a wave number
\begin{equation}
  \label{eq:test_wavenumber}
  k_{qp}
  = \left[ \left(\frac{2\pi}{N_x}\,q\right)^2 + \left(\frac{2\pi}{N_z}\,p\right)^2 \right]^{1/2}  \;.
\end{equation}
Let us define the harmonics as random Gaussian variables with zero mean
\begin{equation}
  \langle \tilde{u}_{qp} \rangle = 0  \;,
\end{equation}
and variance
\begin{equation}
  4 \sum_{k_{qp} < k} \langle |\tilde{u}_{qp}|^2 \rangle
  = 2\pi \int_0^k dk'\,k' P(k')  \;,
\end{equation}
where $P$ is the power spectrum of turbulent velocity pulsations; multipliers $4$ (left-hand side) and $2\pi$ (right-hand side) are needed to take account of turbulence isotropy. We assume that the harmonics' phases are subject to a random uniform distribution in the interval from $0$ to $2\pi$. The resulting radial velocity field is shown in fig.~\ref{fig:test_gaussian_field_vel}.

For each line of sight $l$ we calculate an intensity distribution in the velocity space according to (\ref{eq:intensity_exact}):
\begin{equation}
  \label{eq:test_intensity}
  I(v, l)
  = \sum_{m = 0}^{N_z - 1} \phi[ v - u_{lm}, \sigma_\mathrm{th} ]  \;,
\end{equation}
and a distribution of filtered intensity
\begin{equation}
  I_L(v, l) = \sum_{l'} B_L(l' - l)\,I(v, l)
\end{equation}
(the limits depend on the filter). For each profile $I_L(v, l)$ we calculate the position of the line center:
\begin{equation}
  \overline{v}_L(l) = \int dv\,v\,I_L(v, l)  \;.
\end{equation}

Using the sample of profiles we should calculated the average radial velocity (as we expected, it's value appears to be close to zero)
\begin{equation}
  \langle \overline{v}_L \rangle = \frac{1}{N_x}
    \sum_{l = 0}^{N_x - 1} \overline{v}_L(l)  \;,
\end{equation}
and the unbiased estimate of the line center variance as a function of the filter width:
\begin{equation}
  \label{eq:test_variance}
  U_L = \frac{1}{N_x - 1}
    \sum_{l = 0}^{N_x - 1} \left[ \overline{v}_L(l)
      - \langle \overline{v}_L \rangle \right]^2  \;.
\end{equation}

We have chosen the following parameters of our model. The grid dimensions are $N_x = 5000$, $N_z = 250$. The thermal and turbulent line widths are $\sigma_\mathrm{th} = 25$ and $\sigma_\mathrm{turb} = 0.3 \sigma_\mathrm{th} = 7.5$ respectively. The power spectrum in our 2D model is:
\begin{equation}
  \label{eq:test_gaussian_field_powspec}
  P(k) \propto \left\{
    \begin{aligned}
      & 0 \,,\: & k \leqslant 2\pi/N_z  \;,  \\
      & k^{-\alpha-1} \,,\: & k > 2\pi/N_z  \;.
    \end{aligned}
  \right.
\end{equation}
We assumed the following normalization
\begin{equation}
  2\pi \int_0^\infty dk\,k P(k) = \sigma_\mathrm{turb}^2  \;.
\end{equation}

\begin{figure*}
  \begin{center}
    \includegraphics{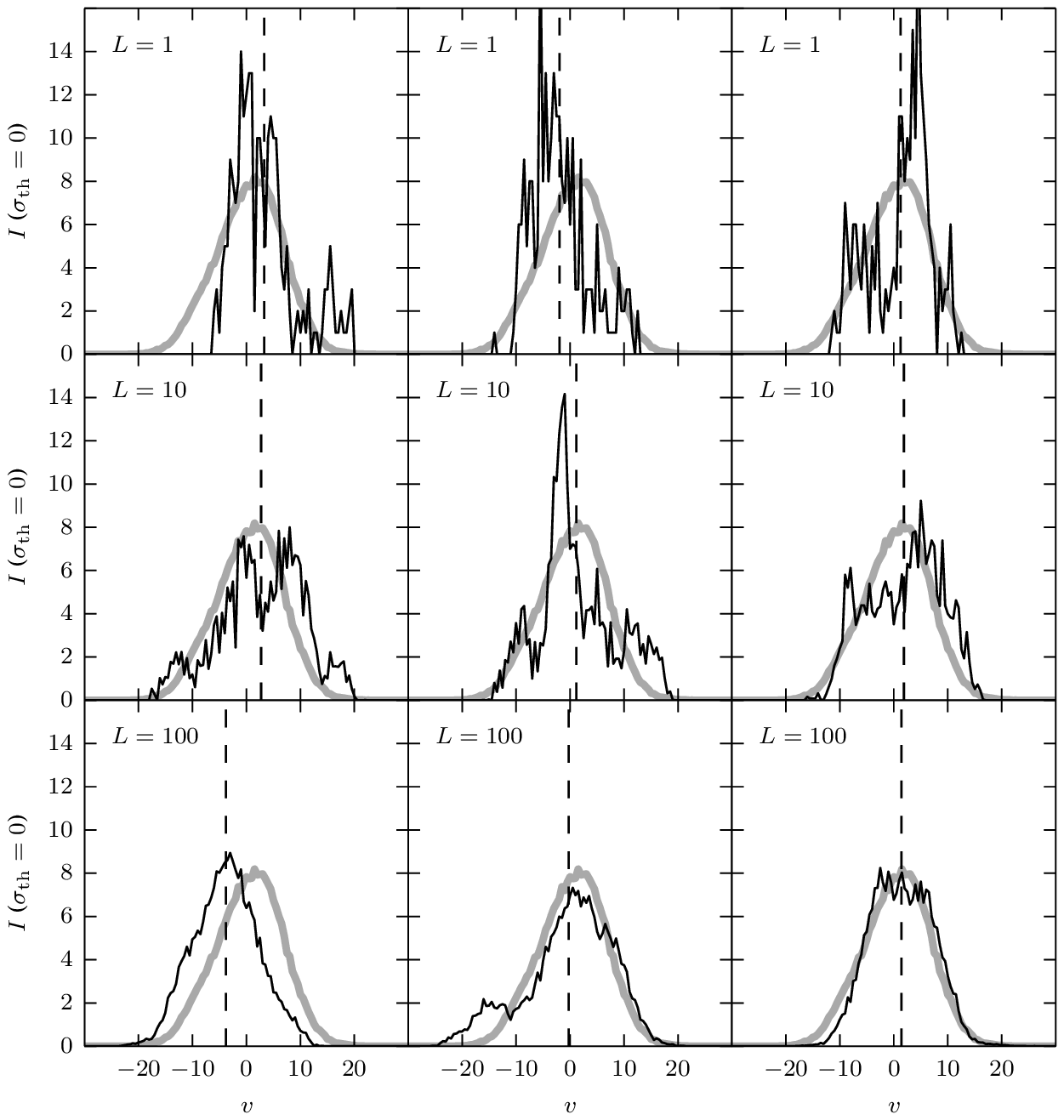}
  \end{center}
  \caption{The distribution of filtered intensity as a function of radial velocity. The thermal broadening was removed from the profiles for visibility. The turbulent medium is approximated by a random Gaussian velocity field. The Gaussian filter is applied. Each column corresponds to a single line of sight. Each row represents the same value of the resolution $L$. Solid gray line depicts a profile averaged over all lines of sight, which is equivalent to $L = 5000$. The dotted line demonstrates the position of the line center.}
  \label{fig:test_gaussian_field_profile}
\end{figure*}

The profile examples free of the contribution of thermal line broadening are shown in fig.~\ref{fig:test_gaussian_field_profile}. One can see that the position of the center of lines varies depending on the resolution $L$. As we expect, when the filter width growth the intensity distribution tends to a profile averaged over all lines of sight, which corresponds to $L = N_x$.

The power spectrum we use (\ref{eq:test_gaussian_field_powspec}) implies the maximum turbulent scale equal to $h_\mathrm{turb} = h = N_z$. Taking into account the other parameters we can derive a theoretical dependence of the line center variance on the filter width:
\begin{equation}
  \label{eq:test_variance_theor}
  U_L = 28.13\,\frac{\alpha - 1}{\alpha}
    \left[ 1 - \left( \frac{L}{250} \right)^\alpha \right]  \;.
\end{equation}

\begin{figure}
  \begin{center}
    \includegraphics[width=0.8\columnwidth]{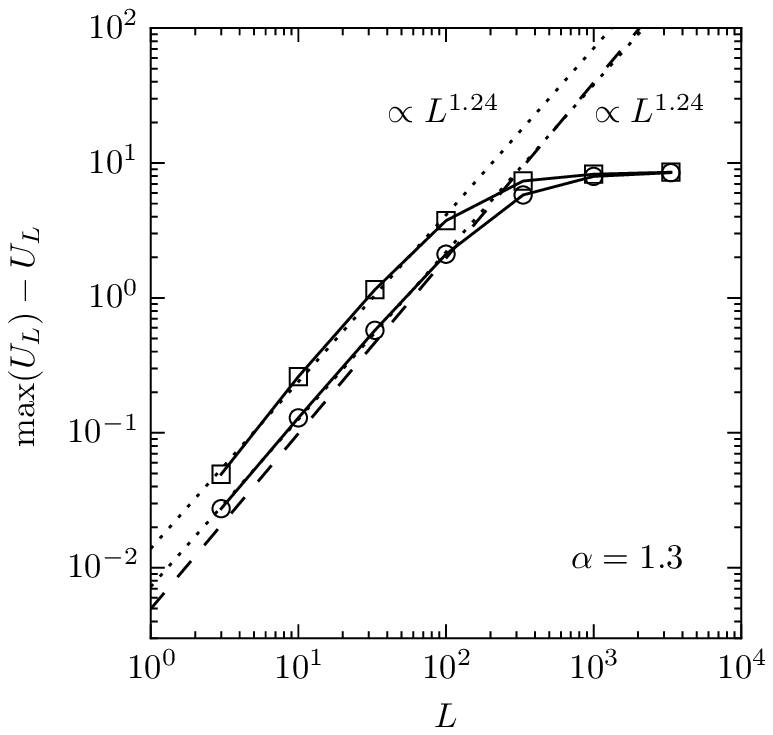}
    \includegraphics[width=0.8\columnwidth]{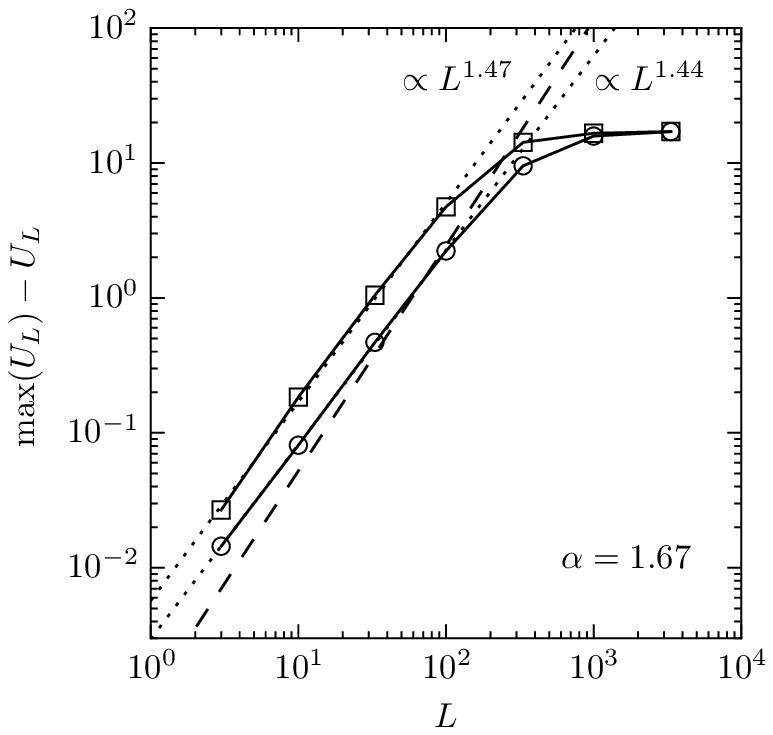}
    \includegraphics[width=0.8\columnwidth]{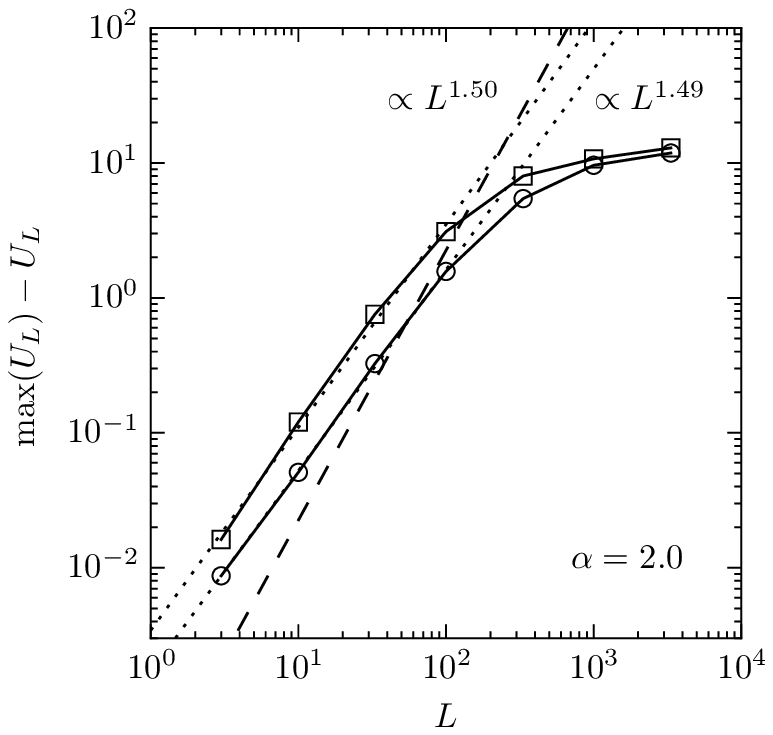}
  \end{center}
  \caption{The variance of line center fluctuations as a function of telescope's resolution in the model where the turbulent medium is approximated by a random Gaussian field of pulsations. We have applied a Gaussian filter (circles) and a box filter (squares). The dashed line depicts the power part of the theoretical function (\ref{eq:test_variance_theor}). Dotted lines are power approximations of the measured variances for $L < 200$, the power indices of the corresponding dotted lines are also shown.}
  \label{fig:test_gaussian_field_variance}
\end{figure}

We have modeled the dependence of $U_L$ for two filters, a Gaussian filter and a box filter with the width $L$, and for various values of exponent of the power spectrum: $\alpha = 1.3, \:1.67, \:2.0$. The obtained results are shown in fig.~\ref{fig:test_gaussian_field_variance}. As one can see that the exponent, estimated from modeled curves, is in a good agreement with the theoretical estimates in the case of $\alpha = 1.3$, and is slightly underestimated for $\alpha = 1.67$ and $\alpha = 2.0$. The reason of this effect is in the fact that we used only the inertial interval of the power spectrum (i.e. short-wavelength), when calculating the theoretical dependence (\ref{eq:variance_approx}), while the long-wavelength part of the power spectrum was only approximated. As the power index $\alpha$ growth, the contribution of the large scale velocity pulsations to $U_L$ growth also. Since this contribution considered approximately the measured curve $\max(U_L) - U_L$ becomes shallower, hence converges more slowly to the power fit (\ref{eq:variance_approx_pow}). We can state that the slope of the modeled function $U_L$ asymptotically approaches the theoretical value in the limit of small filter scales (or high resolution).

\subsection{Rigid-body vortices}

Let us represent the turbulent medium as a random distribution of vortices that fill a 2-D computational domain of $N_x \times N_z$ and do not intersect. Let all the vortices rotate as a rigid-body with the same angular velocity. We assume that the entire volume is filled with isothermal gas, and the radial velocity is equal to zero out of any vortex. The spectrum of vortex sizes is assumed to be power-law.
\begin{figure*}
  \begin{center}
    \includegraphics{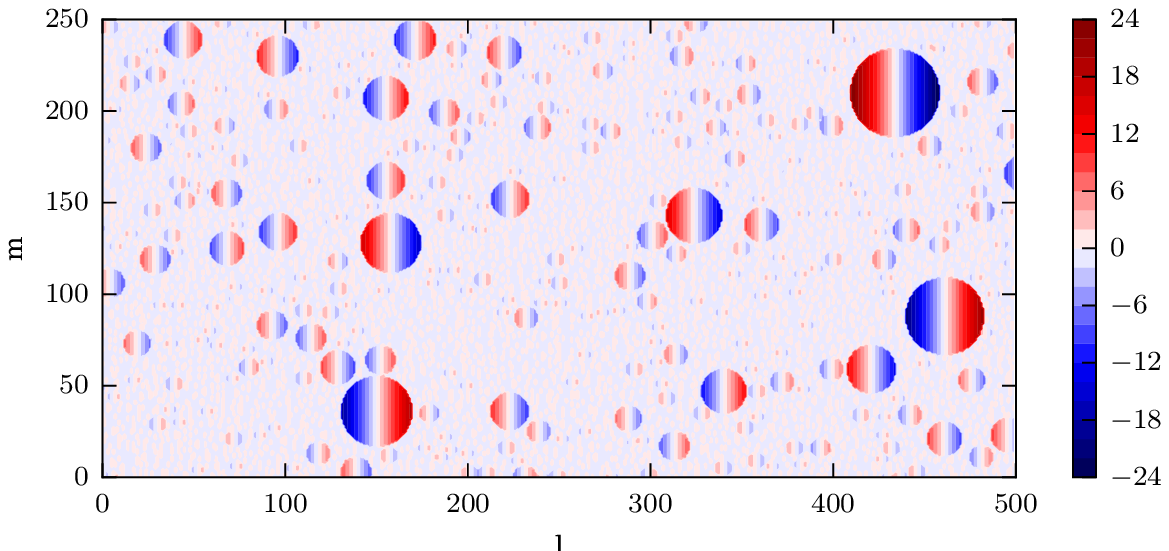}
  \end{center}
  \caption{The field of radial velocities in the test model where the turbulent medium is approximated by a sample of rigid-body rotating vortices.}
  \label{fig:test_solid_verts_vel}
\end{figure*}

Let us first solve the direct problem of the calculation of power spectrum. We should obtain an auto-correlation function of the velocity field:
\begin{equation}
  V_{lm} = \sum_{l' = 0}^{N_x - 1} \sum_{m' = 0}^{N_z - 1} u_{l'm'}\,u_{(l'+l)(m'+m)}
\end{equation}
(we assume that radial velocity is equal to zero outside the computational domain). The Fourier amplitudes of this function -- when depending on a single wave number -- are connected with the power spectrum $P(k)$ of interest via:
\begin{equation}
  \sum_{k_{qp} < k} |\tilde{V}_{qp}|
  \propto \int_0^k dk'\,k' P(k')  \;,
\end{equation}
where
\begin{equation}
  \tilde{V}_{qp} = \sum_{l = 0}^{N_x - 1} \sum_{m = 0}^{N_z - 1}
    \exp\!\left( - i\,\frac{2\pi}{N_x}\,lq - i\,\frac{2\pi}{N_z}\,mp \right)\,V_{lm}  \;,
\end{equation}
and the wave number $k_{qp}$ depends on the wave indexes in accordance with (\ref{eq:test_wavenumber}).

To solve the inverse problem one should transform the radial velocity field into the distributions of intensity, filtered intensity and variance of line center fluctuations according to (\ref{eq:test_intensity})--(\ref{eq:test_variance}).

\begin{figure*}
  \begin{center}
    \includegraphics{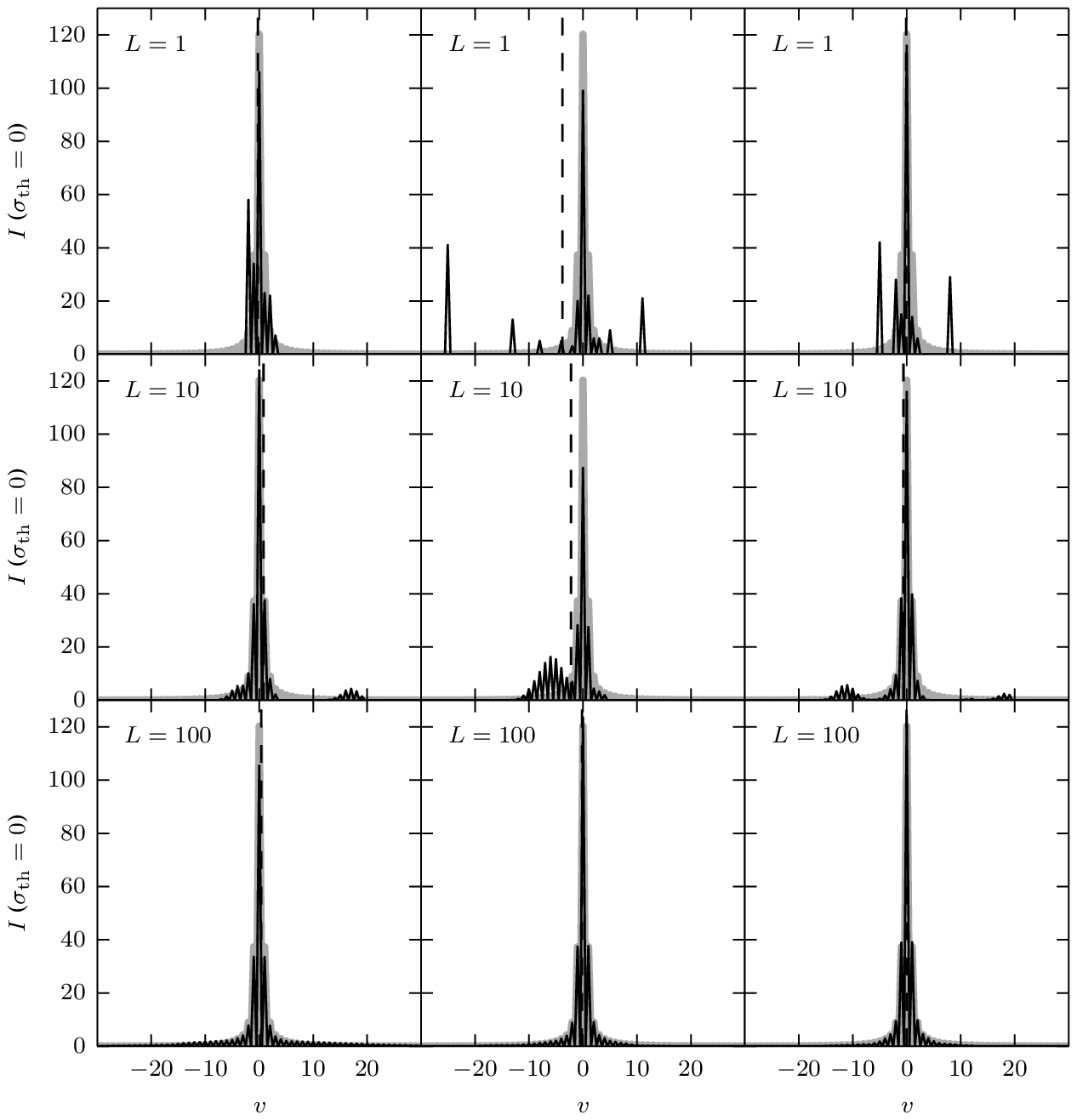}
  \end{center}
  \caption{The distribution of filtered intensity as a function of radial velocity. The thermal broadening was removed from the profiles for visibility. The turbulent medium is represented by a random distribution of rigid-body vortices with a Kolmogorov power spectrum ($\alpha = 1.67$). The denotations are the same as in fig.~\ref{fig:test_gaussian_field_profile}.}
  \label{fig:test_solid_verts_profile}
\end{figure*}

We used the following parameters in our calculations: $N_x = 5000$, $N_z = 250$, $\sigma_\mathrm{th} = 25$. We have calculated two models that differ from each other by the variance of turbulent velocity and the slope of power spectrum. In the first model $\sigma_\mathrm{turb} = 3.9$. The vortices have spatial sizes from $1$ to $100$. The slope of the spectrum of vortex sizes has been chosen to provide a Kolmogorov power spectrum of velocity pulsations with $\alpha = 1.67$. The radial velocity distribution in the computation domain is shown if fig.~\ref{fig:test_solid_verts_vel}. In the second model $\sigma_\mathrm{turb} = 9.8$, $\alpha = 1.9$. The resulting line profiles free of thermal broadening are shown in fig.~\ref{fig:test_solid_verts_profile}. One can see a prominent peak at $v = 0$ in the profiles. It occurs because of the spatial distribution of vortices, namely because we required no intersection of vortices to appear. The distribution of velocity in rigid-body vortices explains the presence of peaks far from the line center. As one can see, all the features in the line profiles are smoothed with the growing filter width, and the resulting profiles tends to its average shape.

\begin{figure}
  \begin{center}
    \includegraphics[width=0.8\columnwidth]{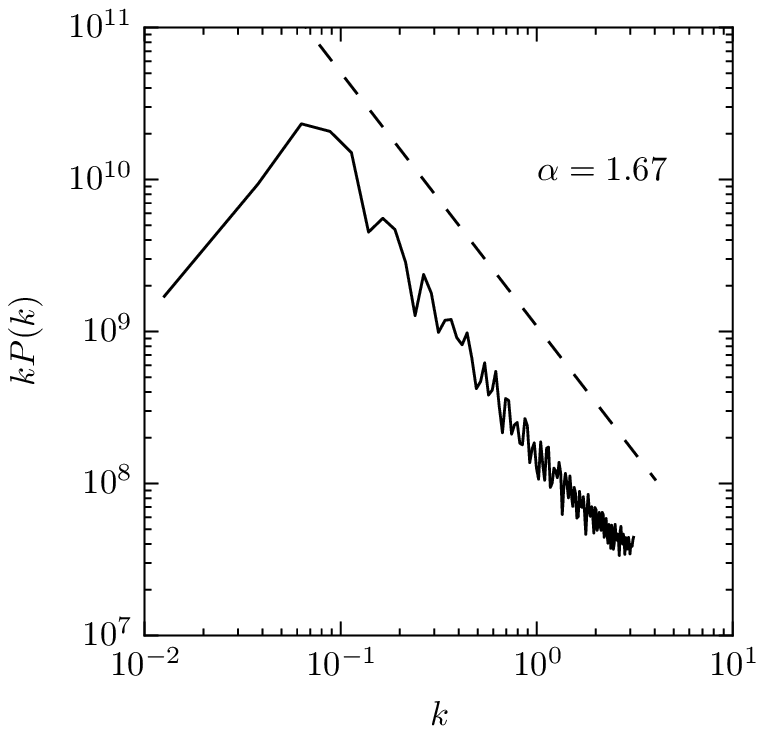} \\
    \includegraphics[width=0.8\columnwidth]{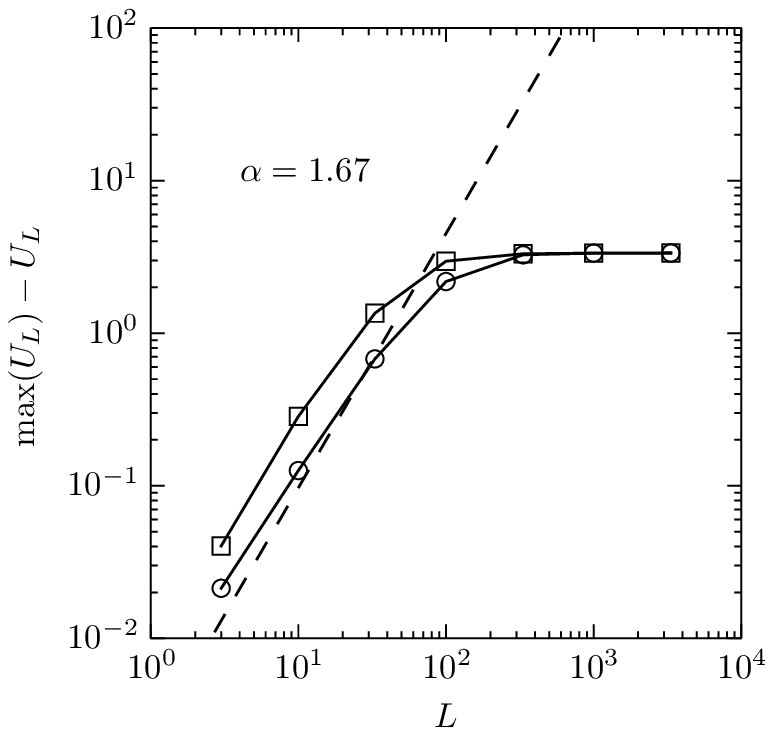}
  \end{center}
  \caption{{\em Top panel}: energy spectrum of turbulent pulsations. The dashed line denotes an approximate slope of the spectrum in the exponential part. {\em Bottom panel}: variance of the line center fluctuations as a function of telescope's resolution. The turbulent medium is represented by a random distribution of rigid-body vortices. We have applied a Gaussian filter (circles) and a rectangular filter (squares). The spectrum of vortex sizes corresponds to  $\alpha = 1.67$.}
  \label{fig:test_solid_verts_1.67_variance}
\end{figure}

\begin{figure}
  \begin{center}
    \includegraphics[width=0.8\columnwidth]{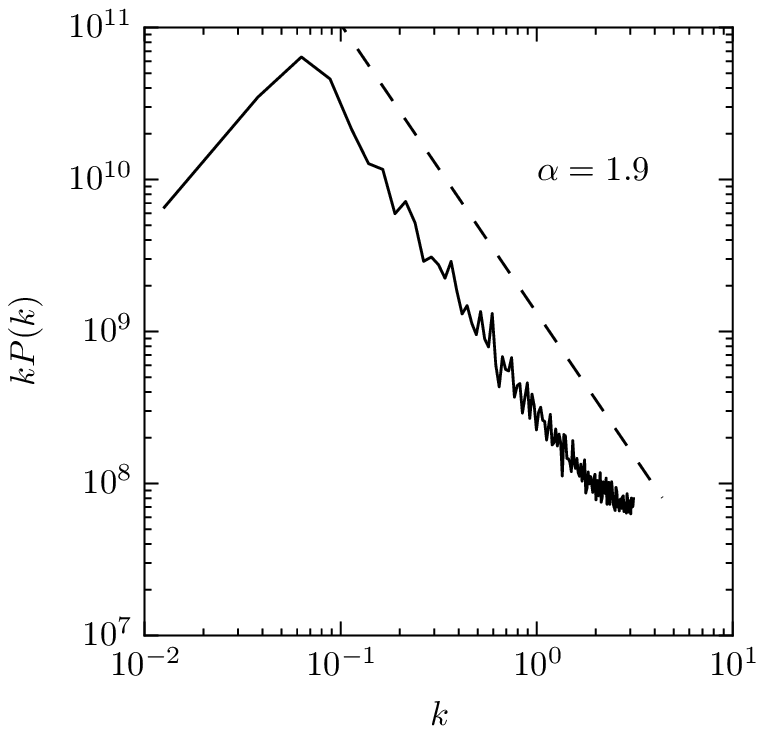} \\
    \includegraphics[width=0.8\columnwidth]{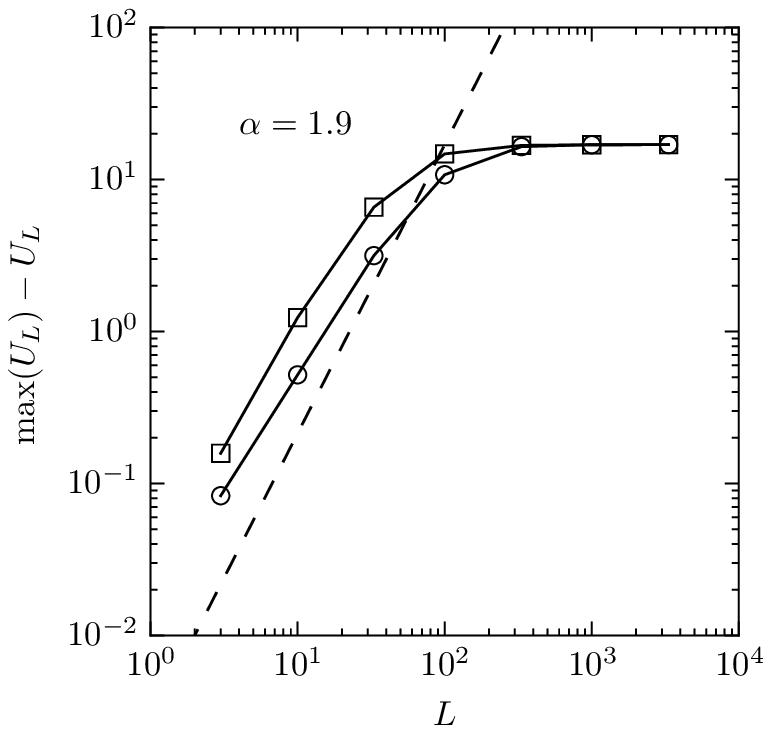}
  \end{center}
  \caption{The same as in fig.~\ref{fig:test_solid_verts_1.67_variance}. The spectrum of vortex sizes corresponds to $\alpha = 1.9$.}
  \label{fig:test_solid_verts_1.9_variance}
\end{figure}

In figures~\ref{fig:test_solid_verts_1.67_variance} and \ref{fig:test_solid_verts_1.9_variance} we show the calculated energy spectra for both the models. The model with the steep power spectrum has resulted in an underestimated value of $\alpha$ as in the case of the Gaussian field of velocity pulsations.

In general, we can argue that the proposed model and method allow us to estimate the slope of the spectrum of turbulence using statistical properties of line profile moments.

\section{Discussion}
\label{sec:discussion}

Above, we presented a method to estimate the slope of the energy spectrum of turbulence in a uniform layer using optically thin emission line profiles observed at different angular resolutions. In fact, the model of a plane-parallel layer is rather simplistic representation for any astrophysical objects of our interest. For instance, protoplanetary disks are expected to have strong gradients of density and temperature, both in vertical and radial directions, see e.g. \citet{Armitage2015arXiv150906382A}. For the inclined protoplanetary disks, there is also a strong line-of-sight regular velocity gradient due to the Keplerian rotation. We expect that parameters of line profiles will strongly depend on angular resolution due to these systematic effects. So the technique of extracting the turbulent spectrum presented above cannot be implemented in its original form since the influence of the systematic physical gradients can be much higher than statistical effects of turbulent fluctuations. However, there is a way to modify the method taking into account the complex structure of the object.

Suppose, that we have a `good' theoretical model of the object that can well prescribe its regular physical structure, i.e. the distribution of density, temperature, regular velocity, concentration of emitting molecules, etc. The parameters of such model can be found via fitting the observed spectra by model spectra. It is important, that the model spectra are calculated taking all the physical effects into account except turbulent fluctuations. Given this model, we can calculate the map of residual spectra i.e. the difference between observed and theoretical spectra. The basic statement of the modified method is that the differences between the observed and model line profiles are supposed to be produced mainly by turbulent fluctuations which are not included in the model. We expect that these differences will depend on the angular resolution of the observed map in the similar way as for the case of uniform layer.

In practice, for the observed map with given angular resolution, we should calculate the variance $U_L$ of the observed line center velocities in respect to the line center positions of the model lines. Note, the line center positions of the model lines will be different over the map since the presence of regular velocity gradients. This procedure should be done for a number of observed maps with different angular resolutions $L$. The dependence of $U_L$ on $L$ will provide us with the information of turbulent spectrum. Since we suggest to calculate variance $U_L$ using all the spectra from the whole map, the implicit assumption of this approach is that the turbulent energy spectrum is the same over the whole object. Of course, this assumption can be over-simplified but we hope that it would give us, at least, the information about the average turbulent spectrum in the object.

The key requirement of the method is to have a set of observed maps with different angular resolutions. These set can be provided not only by different individual observations. In fact, the set of maps can be naturally produced as a result of a single interferometric observation. Indeed, during the reconstruction of the observed spectra map from the observed visibilities we can apply different masks in uv-space and select certain visibilities to be used. The different sets of visibilities will correspond to different angular resolutions of the reconstructed spectra maps.

Another important condition to the developed method is the use of optically thin lines for which the intensity is linearly proportional to the surface density of the layer. In principle, one may try to include the effects of self-absorption and to generalize the method for optically thick lines as well. However, the radiative transfer effects are difficult to tackle and this will make the problem significantly more complex. To our excuse, it is likely possible to select optically thin lines among a number of emission lines available for particular object. For example, if the source is studied in CO lines, its molecular isotopologues C$^{18}$O or $^{13}$CO can be adopted if the emission lines of main isotopologue CO turn out to be optically thick.

The problem to extract turbulent spectrum becomes even more complicated if one takes into account the compressibility of the medium, which is absolutely necessary when considering super-sonic turbulence. In this case the fluctuations of the column density contribute both to the fluctuations of the total line intensity and to the intensity spectral distribution. We plan to address this problem in our future studies.

Among all the astrophysical objects, the protoplanetary disks are likely the most attractive sources to be studied with the described method. There are two general arguments to support this statement. The first argument is that the interferometric observations of protoplanetary disks have become very productive over the last years. For instance, ALMA's longest  baseline is $D = 16$~km, which results in the angular resolution of $0.1''$\footnote{The source of information is the official web page of ALMA https://almascience.eso.org/about-alma/alma-basics}. If one observes a gas-dust disk from the distance of $140$~pc, this angular resolution corresponds to the spatial resolution $\sim 10$~AU. Taking into account the fact that the specific sizes of these disks are up to several hundred AUs, one can see that this spatial resolution is sufficient to study the spatial disk's structure in detail. The maximum spectral resolution of ALMA in terms of velocity is $10$~m/s, which is significantly less than the specific values of the thermal ($\sim 200$~m/s), turbulent ($\sim 100$~m/s) and Keplerian ($\sim 1$~km/s) velocities. This enables studying details of the kinematic structure of gas-dust disks including turbulence.  The high efficiency of ALMA has already bean used to study turbulence in protoplanetary disks. For example, most recent observations of a particular HD~163296 protoplanetary disk and following analysis by \citet{Flaherty2015arXiv151001375F} allowed them to estimate the upper limits of the turbulent motions to be about 3 per cent in respect to thermal line width.

The second argument is based on the fact that validity of the method crucially depends on the assumption that theoretical model successfully reproduce regular (non-stochastic) features of the object and its observed spectra maps. Between the objects where turbulence is important, the protoplanetary disks seem to be most reliably theoretically described. The present models of protoplanetary disks include many physical effects and are able to explain the regular distributions of density, temperature and velocity inside the disks \citep{Armitage2015arXiv150906382A}. With the help of such models it is possible to reconstruct the parameters of the observed disks from the interferometric observations, see e.g. review of \citet{Dutrey2014prpl.conf..317D}. Moreover, the parameter fitting of the disk structure based on interferometric observations have become standard procedure. To our notion, the presented model of deriving the turbulent spectrum can be included as a part of this procedure. However, such implementation of the method as well its application to real disks should be the subject for separate paper. The goal of the current paper was to check if the method could work in principle even for the case of uniform layer.

Now, let us give some notes about energy spectrum of turbulence regardless the method. In the model of the Kolmogorov turbulence the properties of turbulent motion are determined by a single specific parameter $\epsilon$, which is a specific flux of the energy transmitted from large-scale pulsations to small-scale pulsations \citep{Landau1959flme.book.....L}. If we assume this quantity to be constant over the turbulent cascade, then we should not have energy loss within the inertial interval. The energy of pulsations is transmitted from large scales to small scales without loss. The dissipation of energy takes place only on the smallest scale, which is determined by molecular viscosity. Thus, from dimension considerations one can obtain all the main relations for the Kolmogorov turbulence. We should remind the reader that in the frame of this model \citep{Landau1959flme.book.....L} the exponent of the spectrum of turbulence is $\alpha = 5/3$ within the inertial interval, and is $\alpha = 3$ within the dissipative interval.

However, the model of turbulence may be self-similar if one accepts a more general relation
\begin{equation}
  \epsilon = \epsilon_0 \lambda^\beta  \;.
\end{equation}
In this case the energy flux  $\epsilon$ is not constant over the entire turbulent cascade. The exponent $\beta$ characterizes energy loss within the inertial interval during the transition from large to small scales. This energy loss may be caused by, for instance, radiation. $\beta = 0$ corresponds to the Kolmogorov turbulence. If $\beta > 0$ the energy flux decreases from large to small scales, and, oppositely, at $\beta < 0$ it increases from large to small scales. Within this approach from dimension considerations we can easily derive the following relation:
\begin{equation}
  \alpha = \frac{5}{3} + \frac{2}{3} \beta.
\end{equation}
Thus, the case with $\alpha = 3$ corresponds to $\beta = 2$. This means that the energy dissipation due to radiation should take place mostly on large scales and should decrease toward smaller scales.

\section{Conclusions}

In this work we propose a method allowing us to restore the parameters of sub-sonic turbulence in astrophysical disks from the profiles of optically thin emission lines observed at different angular resolutions. The method allows finding the slope of the energy spectrum of turbulence and, in some cases, it allows us to estimate the largest spatial scale of turbulence and the height of the turbulent layer.

Within this method we analyze the fluctuations of intensity in the line profile. The fluctuations are caused by turbulent velocity pulsations in the medium. We show that the finite angular resolution of a telescope influences the statistical properties of the fluctuations in the observed line profile. The effect of the finite resolution is in the fact that the turbulent velocity pulsations on the scales that are below the angular resolution of the telescope are filtered out and do not contribute to the intensity fluctuations.

The efficiency of the method has been demonstrated on numerical models of the turbulent medium. We have calculated two fundamentally different models. In the first model we represented the turbulent medium by a Gaussian random velocity field. The second model was a set of rigidly rotating vortices of various sizes. For both the models the method gives the exponents of the energy spectrum that are in a good agreement with the theoretical value.

The feature of the proposed method is in the fact that in order to use it one needs to have only one sample of 3-D spectroscopic observations, obtained with high angular resolution. Then, by performing spatial filtering in various spectral ranges one can obtain filtered profiles, from which it is possible to recover the parameters of the turbulent medium. The algorithm and applicability conditions of this method may be taken into account as recommendations when planning radio-interferometric observations.

We tested the proposed method of the recovering the turbulent power spectrum only for the idealized models. Its application to real protoplanetary disks requires high-quality observational data which became available with with the start of ALMA observatory. In the following papers we plan to adapt and implement this method to real protoplanetary disks.

\section*{Acknowledgements}

D.~V.~Bisikalo and A.~G.~Zhilkin are supported by the Russian Science Foundation (project 15-12-30038).  E.~P.~Kurbatov is supported by the Russian Foundation for Basic Research (project 14-29-06059). Ya.~N.~Pavlyuchenkov is supported by the Russian Foundation for Basic Research (projects 14-02-00719 and 16-02-00834).



\bibliographystyle{mnras}
\bibliography{paper}

\begin{thebibliography}{}
\makeatletter
\relax
\def\mn@urlcharsother{\let\do\@makeother \do\$\do\&\do\#\do\^\do\_\do\%\do\~}
\def\mn@doi{\begingroup\mn@urlcharsother \@ifnextchar [ {\mn@doi@}
  {\mn@doi@[]}}
\def\mn@doi@[#1]#2{\def\@tempa{#1}\ifx\@tempa\@empty \href
  {http://dx.doi.org/#2} {doi:#2}\else \href {http://dx.doi.org/#2} {#1}\fi
  \endgroup}
\def\mn@eprint#1#2{\mn@eprint@#1:#2::\@nil}
\def\mn@eprint@arXiv#1{\href {http://arxiv.org/abs/#1} {{\tt arXiv:#1}}}
\def\mn@eprint@dblp#1{\href {http://dblp.uni-trier.de/rec/bibtex/#1.xml}
  {dblp:#1}}
\def\mn@eprint@#1:#2:#3:#4\@nil{\def\@tempa {#1}\def\@tempb {#2}\def\@tempc
  {#3}\ifx \@tempc \@empty \let \@tempc \@tempb \let \@tempb \@tempa \fi \ifx
  \@tempb \@empty \def\@tempb {arXiv}\fi \@ifundefined
  {mn@eprint@\@tempb}{\@tempb:\@tempc}{\expandafter \expandafter \csname
  mn@eprint@\@tempb\endcsname \expandafter{\@tempc}}}

\bibitem[\protect\citeauthoryear{{Armitage}}{{Armitage}}{2015}]{Armitage2015arXiv150906382A}
{Armitage} P.~J.,  2015, preprint, \href
  {http://adsabs.harvard.edu/abs/2015arXiv150906382A} {} (\mn@eprint {arXiv}
  {1509.06382})

\bibitem[\protect\citeauthoryear{{Bisnovatyi-Kogan} \&
  {Blinnikov}}{{Bisnovatyi-Kogan} \&
  {Blinnikov}}{1976}]{Bisnovatyi-Kogan1976PAZh....2..489B}
{Bisnovatyi-Kogan} G.~S.,  {Blinnikov} S.~I.,  1976, Pisma v Astronomicheskii
  Zhurnal, \href {http://adsabs.harvard.edu/abs/1976PAZh....2..489B} {2, 489}

\bibitem[\protect\citeauthoryear{{Blandford} \& {Payne}}{{Blandford} \&
  {Payne}}{1982}]{Blandford1982MNRAS.199..883B}
{Blandford} R.~D.,  {Payne} D.~G.,  1982, \mnras, \href
  {http://adsabs.harvard.edu/abs/1982MNRAS.199..883B} {199, 883}

\bibitem[\protect\citeauthoryear{{Cao} \& {Spruit}}{{Cao} \&
  {Spruit}}{2002}]{Cao2002A&A...385..289C}
{Cao} X.,  {Spruit} H.~C.,  2002, \mn@doi [\aap] {10.1051/0004-6361:20011818},
  \href {http://adsabs.harvard.edu/abs/2002A%26A...385..289C} {385, 289}

\bibitem[\protect\citeauthoryear{{Dartois}, {Dutrey}  \&
  {Guilloteau}}{{Dartois} et~al.}{2003}]{Dartois2003A&A...399..773D}
{Dartois} E.,  {Dutrey} A.,   {Guilloteau} S.,  2003, \mn@doi [\aap]
  {10.1051/0004-6361:20021638}, \href
  {http://adsabs.harvard.edu/abs/2003A%26A...399..773D} {399, 773}

\bibitem[\protect\citeauthoryear{{Dgani}, {Livio}  \& {Regev}}{{Dgani}
  et~al.}{1994}]{Dgani1994ApJ...436..270D}
{Dgani} R.,  {Livio} M.,   {Regev} O.,  1994, \mn@doi [\apj] {10.1086/174901},
  \href {http://adsabs.harvard.edu/abs/1994ApJ...436..270D} {436, 270}

\bibitem[\protect\citeauthoryear{{Dutrey} et~al.,}{{Dutrey}
  et~al.}{2014}]{Dutrey2014prpl.conf..317D}
{Dutrey} A.,  et~al., 2014, \mn@doi [Protostars and Planets VI]
  {10.2458/azu_uapress_9780816531240-ch014}, \href
  {http://adsabs.harvard.edu/abs/2014prpl.conf..317D} {pp 317--338}

\bibitem[\protect\citeauthoryear{{Flaherty}, {Hughes}, {Rosenfeld}, {Andrews},
  {Chiang}, {Simon}, {Kerzner}  \& {Wilner}}{{Flaherty}
  et~al.}{2015}]{Flaherty2015arXiv151001375F}
{Flaherty} K.~M.,  {Hughes} A.~M.,  {Rosenfeld} K.~A.,  {Andrews} S.~M.,
  {Chiang} E.,  {Simon} J.~B.,  {Kerzner} S.,   {Wilner} D.~J.,  2015,
  preprint, \href {http://adsabs.harvard.edu/abs/2015arXiv151001375F} {}
  (\mn@eprint {arXiv} {1510.01375})

\bibitem[\protect\citeauthoryear{{Guilloteau} \& {Dutrey}}{{Guilloteau} \&
  {Dutrey}}{1998}]{Guilloteau1998A&A...339..467G}
{Guilloteau} S.,  {Dutrey} A.,  1998, \aap, \href
  {http://adsabs.harvard.edu/abs/1998A%26A...339..467G} {339, 467}

\bibitem[\protect\citeauthoryear{{Guilloteau}, {Dutrey}, {Wakelam}, {Hersant},
  {Semenov}, {Chapillon}, {Henning}  \& {Pi{\'e}tu}}{{Guilloteau}
  et~al.}{2012}]{Guilloteau2012A&A...548A..70G}
{Guilloteau} S.,  {Dutrey} A.,  {Wakelam} V.,  {Hersant} F.,  {Semenov} D.,
  {Chapillon} E.,  {Henning} T.,   {Pi{\'e}tu} V.,  2012, \mn@doi [\aap]
  {10.1051/0004-6361/201220331}, \href
  {http://adsabs.harvard.edu/abs/2012A%26A...548A..70G} {548, A70}

\bibitem[\protect\citeauthoryear{{Hartmann}, {Calvet}, {Gullbring}  \&
  {D'Alessio}}{{Hartmann} et~al.}{1998}]{Hartmann1998ApJ...495..385H}
{Hartmann} L.,  {Calvet} N.,  {Gullbring} E.,   {D'Alessio} P.,  1998, \mn@doi
  [\apj] {10.1086/305277}, \href
  {http://adsabs.harvard.edu/abs/1998ApJ...495..385H} {495, 385}

\bibitem[\protect\citeauthoryear{{Landau} \& {Lifshitz}}{{Landau} \&
  {Lifshitz}}{1959}]{Landau1959flme.book.....L}
{Landau} L.~D.,  {Lifshitz} E.~M.,  1959, {Fluid mechanics}.
Course of theoretical physics, Oxford: Pergamon Press, 1959

\bibitem[\protect\citeauthoryear{{Lazarian} \& {Pogosyan}}{{Lazarian} \&
  {Pogosyan}}{2000}]{Lazarian2000ApJ...537..720L}
{Lazarian} A.,  {Pogosyan} D.,  2000, \mn@doi [\apj] {10.1086/309040}, \href
  {http://adsabs.harvard.edu/abs/2000ApJ...537..720L} {537, 720}

\bibitem[\protect\citeauthoryear{{Lazarian} \& {Pogosyan}}{{Lazarian} \&
  {Pogosyan}}{2004}]{Lazarian2004ApJ...616..943L}
{Lazarian} A.,  {Pogosyan} D.,  2004, \mn@doi [\apj] {10.1086/422462}, \href
  {http://adsabs.harvard.edu/abs/2004ApJ...616..943L} {616, 943}

\bibitem[\protect\citeauthoryear{{Lazarian} \& {Pogosyan}}{{Lazarian} \&
  {Pogosyan}}{2006}]{Lazarian2006ApJ...652.1348L}
{Lazarian} A.,  {Pogosyan} D.,  2006, \mn@doi [\apj] {10.1086/508012}, \href
  {http://adsabs.harvard.edu/abs/2006ApJ...652.1348L} {652, 1348}

\bibitem[\protect\citeauthoryear{{Lin} \& {Papaloizou}}{{Lin} \&
  {Papaloizou}}{1979}]{Lin1979MNRAS.186..799L}
{Lin} D.~N.~C.,  {Papaloizou} J.,  1979, \mnras, \href
  {http://adsabs.harvard.edu/abs/1979MNRAS.186..799L} {186, 799}

\bibitem[\protect\citeauthoryear{{Lin} \& {Papaloizou}}{{Lin} \&
  {Papaloizou}}{1980}]{Lin1980MNRAS.191...37L}
{Lin} D.~N.~C.,  {Papaloizou} J.,  1980, \mnras, \href
  {http://adsabs.harvard.edu/abs/1980MNRAS.191...37L} {191, 37}

\bibitem[\protect\citeauthoryear{{Lin}, {Papaloizou}  \& {Savonije}}{{Lin}
  et~al.}{1990}]{Lin1990ApJ...365..748L}
{Lin} D.~N.~C.,  {Papaloizou} J.~C.~B.,   {Savonije} G.~J.,  1990, \mn@doi
  [\apj] {10.1086/169528}, \href
  {http://adsabs.harvard.edu/abs/1990ApJ...365..748L} {365, 748}

\bibitem[\protect\citeauthoryear{{Lubow}}{{Lubow}}{1981}]{Lubow1981ApJ...245..274L}
{Lubow} S.~H.,  1981, \mn@doi [\apj] {10.1086/158808}, \href
  {http://adsabs.harvard.edu/abs/1981ApJ...245..274L} {245, 274}

\bibitem[\protect\citeauthoryear{{Lubow}, {Papaloizou}  \& {Pringle}}{{Lubow}
  et~al.}{1994}]{Lubow1994MNRAS.268.1010L}
{Lubow} S.~H.,  {Papaloizou} J.~C.~B.,   {Pringle} J.~E.,  1994, \mnras, \href
  {http://adsabs.harvard.edu/abs/1994MNRAS.268.1010L} {268, 1010}

\bibitem[\protect\citeauthoryear{{Lynden-Bell}}{{Lynden-Bell}}{2003}]{Lynden-Bell2003MNRAS.341.1360L}
{Lynden-Bell} D.,  2003, \mn@doi [\mnras] {10.1046/j.1365-8711.2003.06506.x},
  \href {http://adsabs.harvard.edu/abs/2003MNRAS.341.1360L} {341, 1360}

\bibitem[\protect\citeauthoryear{{Lynden-Bell} \& {Pringle}}{{Lynden-Bell} \&
  {Pringle}}{1974}]{Lynden-Bell1974MNRAS.168..603L}
{Lynden-Bell} D.,  {Pringle} J.~E.,  1974, \mnras, \href
  {http://adsabs.harvard.edu/abs/1974MNRAS.168..603L} {168, 603}

\bibitem[\protect\citeauthoryear{{Michel}}{{Michel}}{1984}]{Michel1984ApJ...279..807M}
{Michel} F.~C.,  1984, \mn@doi [\apj] {10.1086/161950}, \href
  {http://adsabs.harvard.edu/abs/1984ApJ...279..807M} {279, 807}

\bibitem[\protect\citeauthoryear{{Paczynski}}{{Paczynski}}{1976}]{Paczynski1976ComAp...6...95P}
{Paczynski} B.,  1976, Comments on Astrophysics, \href
  {http://adsabs.harvard.edu/abs/1976ComAp...6...95P} {6, 95}

\bibitem[\protect\citeauthoryear{{Papaloizou} \& {Pringle}}{{Papaloizou} \&
  {Pringle}}{1977}]{Papaloizou1977MNRAS.181..441P}
{Papaloizou} J.,  {Pringle} J.~E.,  1977, \mnras, \href
  {http://adsabs.harvard.edu/abs/1977MNRAS.181..441P} {181, 441}

\bibitem[\protect\citeauthoryear{{Sawada}, {Matsuda}  \& {Hachisu}}{{Sawada}
  et~al.}{1986}]{Sawada1986MNRAS.219...75S}
{Sawada} K.,  {Matsuda} T.,   {Hachisu} I.,  1986, \mnras, \href
  {http://adsabs.harvard.edu/abs/1986MNRAS.219...75S} {219, 75}

\bibitem[\protect\citeauthoryear{{Shakura}}{{Shakura}}{1972}]{Shakura1972AZh....49..921S}
{Shakura} N.~I.,  1972, Astron. Zh., \href
  {http://adsabs.harvard.edu/abs/1972AZh....49..921S} {49, 921}

\bibitem[\protect\citeauthoryear{Shakura}{Shakura}{2014}]{Shakura2014PhyU..184..445S}
Shakura N.~I.,  2014, \mn@doi [Physics-Uspekhi]
  {10.3367/UFNe.0184.201404h.0445}, 57, 407

\bibitem[\protect\citeauthoryear{{Shakura} \& {Sunyaev}}{{Shakura} \&
  {Sunyaev}}{1973}]{Shakura1973A&A....24..337S}
{Shakura} N.~I.,  {Sunyaev} R.~A.,  1973, \aap, \href
  {http://adsabs.harvard.edu/abs/1973A%26A....24..337S} {24, 337}

\bibitem[\protect\citeauthoryear{{Shakura}, {Sunyaev}  \&
  {Zilitinkevich}}{{Shakura} et~al.}{1978}]{Shakura1978A&A....62..179S}
{Shakura} N.~I.,  {Sunyaev} R.~A.,   {Zilitinkevich} S.~S.,  1978, \aap, \href
  {http://adsabs.harvard.edu/abs/1978A%26A....62..179S} {62, 179}

\bibitem[\protect\citeauthoryear{{Spruit}}{{Spruit}}{1987}]{Spruit1987A&A...184..173S}
{Spruit} H.~C.,  1987, \aap, \href
  {http://adsabs.harvard.edu/abs/1987A%26A...184..173S} {184, 173}

\bibitem[\protect\citeauthoryear{{Syer} \& {Narayan}}{{Syer} \&
  {Narayan}}{1993}]{Syer1993MNRAS.262..749S}
{Syer} D.,  {Narayan} R.,  1993, \mnras, \href
  {http://adsabs.harvard.edu/abs/1993MNRAS.262..749S} {262, 749}

\bibitem[\protect\citeauthoryear{{Vishniac} \& {Diamond}}{{Vishniac} \&
  {Diamond}}{1989}]{Vishniac1989ApJ...347..435V}
{Vishniac} E.~T.,  {Diamond} P.,  1989, \mn@doi [\apj] {10.1086/168131}, \href
  {http://adsabs.harvard.edu/abs/1989ApJ...347..435V} {347, 435}

\bibitem[\protect\citeauthoryear{{Vorobyov}, {Pavlyuchenkov}  \&
  {Trinkl}}{{Vorobyov} et~al.}{2014}]{Vorobyov2014ARep...58..522V}
{Vorobyov} E.~I.,  {Pavlyuchenkov} Y.~N.,   {Trinkl} P.,  2014, \mn@doi
  [Astronomy Reports] {10.1134/S1063772914080083}, \href
  {http://adsabs.harvard.edu/abs/2014ARep...58..522V} {58, 522}

\makeatother
\end{thebibliography}



\appendix

\section{Formation of line profile}

The equation of radiation transfer at a frequency $\nu$ is:
\begin{equation}
  \label{eq:radiative_transfer_app}
  \Dfrac{I_\nu}{z} = j_\nu  \;,
\end{equation}
where $z$ is the coordinate along the line of sight; $j_\nu$ is the emissivity at the frequency $\nu$.

The particle velocity obeys the Maxwellian distribution with the mean radial velocity $u$ and the variance along a single direction $\sigma_\mathrm{th}^2$:
\begin{equation}
  \label{eq:maxwellian_distribution_app}
  \phi(v - u, \sigma_\mathrm{th})\,dv
  = \frac{1}{\sqrt{2\pi \sigma_\mathrm{th}^2}}\,
    \operatorname{exp}\left[ - \frac{(v - u)^2}{2 \sigma_\mathrm{th}^2} \right]\,dv  \;.
\end{equation}
The source in the equation of radiation transfer is defined as a convolution of an elementary profile of particle emission and the Maxwellian distribution of for the velocity:
\begin{align}
  j_\nu
  &{}= \int dv\,\phi(v - u, \sigma)\,\varepsilon n\,
    \delta_\mathrm{D}\!\left[ \nu - \nu_0 \left( 1 - \frac{v}{c} \right) \right]  \\
  &{}= \varepsilon n \phi\big[ \nu - \nu_0 (1 - u/c), \Delta_\mathrm{th} \big]  \;,
\end{align}
where $n$ is the number density of emitting particles; $\delta_\mathrm{D}[\dots]$ is the Dirac delta function; $\Delta_\mathrm{th} = \sigma_\mathrm{th} \nu_0/c$ is the thermal line width; $c$ is the speed of light.

The number density and radial velocity depend on the spatial coordinates and time, i.e. different volumes in the medium may have different values of the bulk velocity and number density. We assume that the temperature is constant over the entire volume. The total intensity along a line of sight, given by the radius-vector $\vecb{R} = (x, y)$ in the image plane is
\begin{equation}
  I(\nu, \vecb{R}) =
    \varepsilon \int dz\,n(\vecb{r})\,
    \phi\big[ \nu - \nu_0 (1 - u(\vecb{r})/c), \Delta_\mathrm{th}(\vecb{r}) \big]  \;,
\end{equation}
where $\vecb{r} = (x, y, z)$. Hereafter, for the sake of convenience we express the intensity as a function of the radial velocity, given that the frequency and radial velocity are connected via the Doppler shift:
\begin{equation}
  \nu = \nu_0 \left( 1 - \frac{v}{c} \right)  \;.
\end{equation}
The same expression for a line profile, associated with a line of sight $\vecb{R}$, is
\begin{equation}
  \label{eq:intensity_exact_app}
  I(v, \vecb{R})
  = \varepsilon \int dz\,n(\vecb{r})\,\phi[ v - u(\vecb{r}), \sigma_\mathrm{th}(\vecb{r}) ]  \;.
\end{equation}
In the expression above the line profile is a weighted sum of Gaussian profiles. The center and the thermal width of each profile may vary in the volume of the medium, and the number density acting as the weight.

Total intensity integrated over all frequencies is:
\begin{equation}
  I(\vecb{R}) = \int dv\,I(v, \vecb{R}) = \varepsilon N(\vecb{R})  \;,
\end{equation}
where
\begin{equation}
  N(\vecb{R}) = \int dz\,n(\vecb{r})
\end{equation}
is the column number density.

The center frequency of a line may differ from the laboratory rest of line $\nu_0$, since the mean velocity in the medium along the line of sight may differ from zero:
\begin{equation}
  \label{eq:mean_frequency_app}
  \overline{v}(\vecb{R})
  = \frac{1}{I(\vecb{R})} \int dv\,v I(v, \vecb{R})
  = \int dz\,\frac{n(\vecb{r})}{N(\vecb{R})}\,u(\vecb{r})  \;.
\end{equation}
The line width may be found as the second moment of the intensity distribution:
\begin{equation}
  \overline{[v - \overline{v}(\vecb{R})]^2}
  = \frac{1}{I(\vecb{R})} \int dv\,[v - \overline{v}(\vecb{R})]^2 I(v, \vecb{R})  \;.
\end{equation}
By using equations (\ref{eq:intensity_exact_app}) and (\ref{eq:mean_frequency_app}) we can easily show that
\begin{equation}
  \label{eq:total_line_width_square_app}
  \overline{[v - \overline{v}(\vecb{R})]^2}
  = \int dz\,\frac{n(\vecb{r})}{N(\vecb{R})}\,[\sigma_\mathrm{th}^2(\vecb{r}) + u^2(\vecb{r})]
    - \overline{v}^2(\vecb{R})  \;.
\end{equation}

\section{Statistics of fluctuations}

The intensity at a given frequency is a random variable for each line of sight. It depends both on the fluctuations of the radial velocity of particles and the fluctuations of density. If the spectrum of turbulence is of Kolmogorov type, the density fluctuations play insignificant role in the formation of an emission line \cite{Lazarian2006ApJ...652.1348L}, and one can neglect them. This is also supported by the fact that, strictly speaking, the Kolmogorov theory is correct only for incompressible medium. Below we allow the number density to change along the line of sight only, i.e. $n = n(z)$.

We denote the averaging procedure over the ensemble of realizations of a random velocity field as $\langle \cdot \rangle$. Without loss of generality we assume that the ensemble average of the radial velocity is equal to zero everywhere $\langle u(\vecb{r}) \rangle \equiv 0$. We also assume that the field of velocity pulsations is statistically uniform and isotropic. Then we can argue that all the odd moments of velocity are equal to zero $\langle [ u(\vecb{r}) ]^{2n+1} \rangle \equiv 0$ where $n$ is a positive integer. The second moment is the variance of turbulent velocity:
\begin{equation}
  \sigma_\mathrm{turb}^2 = \langle u^2(\vecb{r}) \rangle  \;.
\end{equation}

Let us introduce a filtered line profile as\footnote{%
Hereafter we omit the explicit dependence on $\vecb{R}$, implying that different realizations of the line profile $I(v)$ correspond to different lines of sight.}
\begin{equation}
  \label{eq:intensity_filtered_app}
  I_L(v)
  = \varepsilon N \int d^3r\,G_L(\vecb{r})
    \,\phi[ v - u(\vecb{r}), \sigma_\mathrm{th}(\vecb{r}) ]  \;,
\end{equation}
where the column number density $N$ is constant, due to the assumption $n = n(z)$; $L$ is the spatial filtering scale. In this expression we simultaneously filter and integrate along the line of sight by convolution the integrand with a kernel $G(\vecb{r})$:
\begin{equation}
  G_L(\vecb{r})
  = B_L(\vecb{R})\,\frac{n(z)}{N}  \;.
\end{equation}
The convolution kernel must satisfy a normalization $\int d^3r\,G_L(\vecb{r}) = 1$.

The central velocity of the filtered profile can be found in the same way as in (\ref{eq:mean_frequency_app}):
\begin{equation}
  \overline{v}_L
  = \frac{1}{Q N} \int dv\,v I_L(v)
  = \int d^3r\,G_L(\vecb{r})\,u(\vecb{r})  \;.
\end{equation}
Its ensemble mean is equal to zero, and the variance is given by the spatial pair correlation function of radial velocity:
\begin{equation}
  U_L = \left\langle \overline{v}_L^2 \right\rangle
  = \int d^3r\,G_L(\vecb{r}) \int d^3r'\,G_L(\vecb{r}')
      \,\langle u(\vecb{r})\,u(\vecb{r}') \rangle  \;.
\end{equation}
In the Fourier\footnote{%
We define the one-dimensional Fourier transform as:
$\tilde{f}(k) = (2\pi)^{-1} \int dx\,e^{i kx}\,f(x)$,
$f(x) = \int dk\,e^{-i kx}\,\tilde{f}(k)$.
}
space this expression is
\begin{equation}
  \label{eq:fluctuations_level_general_app}
  U_L
  = (2\pi)^6 \int d^3k\,\tilde{G}_L(\vecb{k})
    \int d^3k'\,\tilde{G}_L^\ast(\vecb{k}')
    \,\langle \tilde{u}(\vecb{k})\,\tilde{u}^\ast(\vecb{k}') \rangle  \;.
\end{equation}
Let us assume that the amplitude of the Fourier harmonics are not mutually correlated and are isotropically distributed. In this case the spatial correlation function is determined by the power spectrum,
\begin{equation}
  \langle \tilde{u}(\vecb{k})\,\tilde{u}^\ast(\vecb{k}') \rangle =
  \delta_\mathrm{D}(\vecb{k} - \vecb{k}')\,P(k) \;,
\end{equation}
and (\ref{eq:fluctuations_level_general_app}) can be rewritten as
\begin{equation}
  \label{eq:fluctuations_level_app}
  U_L = (2\pi)^6 \int d^3k\,|\tilde{G}_L(\vecb{k})|^2 P(k)  \;.
\end{equation}

Before we start calculating this value let us consider the power spectrum of the turbulent velocity field. Within our approach we can pick out three specific spatial scales: the characteristic disk height $h$ which is determined by the distribution $n(z)$, e.g. via it's second momentum; the largest turbulent scale $h_\mathrm{turb}$; and the scale at which turbulence dissipates $\lambda$, here $h \gtrsim h_\mathrm{turb} \gg \lambda$. Let us assume the following properties of the turbulence:
\begin{itemize}
  \item in the inertial interval ($h_\mathrm{turb}^{-1} \ll k \ll \lambda^{-1}$) the spectrum decreases according to the power law,
    \begin{equation}
    \label{eq:power_spectrum_app}
    P(k) \equiv \sigma_\mathrm{turb}^2 A k^{-\alpha-2}  \;,
  \end{equation}
  where $A, \alpha > 0$;
  \item on scales larger than the maximal turbulent scale and smaller than the disk height the spectrum decreases with increasing $k^{-1}$;
  \item on scales larger than the disk height the power spectrum quickly approaches zero
\end{itemize}
The specific energy of turbulent pulsations on a certain scale $k^{-1}$ can be estimated as $k^3 P(k) \propto k^{1-\alpha}$. Thus, for $\alpha > 1$ the most energetic pulsations are those on larger scales. Hereafter we assume that this is correct in our case.

The power spectrum is normalized by the velocity variance as follows
\begin{equation}
  \sigma_\mathrm{turb}^2 \equiv \int d^3k\,P(k) = 4\pi \int_0^\infty dk\,k^2 P(k)  \;.
\end{equation}
Let us separate the integration over the inertial interval assuming $\lambda = 0$, since this limit does not contribute much to the normalization:
\begin{align}
  \sigma_\mathrm{turb}^2
  &{}= C \sigma_\mathrm{turb}^2
    + 4\pi \sigma_\mathrm{turb}^2 A \int_{h_\mathrm{turb}^{-1}}^\infty dk\,k^{-\alpha}  \\
  &{}= \sigma_\mathrm{turb}^2 \left[ C
    + \frac{4\pi A}{\alpha-1}\,h_\mathrm{turb}^{\alpha-1} \right]  \;,
  \label{eq:turbulent_velocity_variance_app}
\end{align}
where $C$ is a positive constant determined by integrating over the long-wavelength part of the spectrum. According to our previous considerations we can argue that the second term in the right-hand side of (\ref{eq:turbulent_velocity_variance_app}) is the main contributor into the turbulent velocity variance, while $C$ is close to zero.

We can write the convolution kernel $G(\vecb{r})$ in the Fourier space as
\begin{equation}
  \tilde{G}_L(\vecb{k}) = \tilde{B}_L(k_\perp)\,\frac{\tilde{n}(k_\parallel)}{N}  \;,
\end{equation}
where $k_\perp$ and $k_\parallel$ are the wave numbers in the image plane and along the line of sight, respectively. For the filtering we can use either a Gaussian function with the specific width $L$ as,
\begin{equation}
  B_L(\vecb{R}) = \frac{1}{2\pi L^2}\,\operatorname{exp}\left( -\frac{R^2}{2 L^2} \right)  \;,
\end{equation}
whose Fourier image is
\begin{equation}
  (2\pi)^2 \tilde{B}_L(k_\perp) =
  \operatorname{exp}\left( -\frac{1}{2}\,k_\perp^2 L^2 \right)  \;,
\end{equation}
or a step-function of width $L$:
\begin{equation}
  B_L(\vecb{R}) = \frac{1}{\pi L^2}\,\theta\!\left( 1 - \frac{|R|}{L} \right)  \;,
\end{equation}
whose Fourier image is given by the Bessel function
\begin{equation}
  (2\pi)^2 \tilde{B}_L(k_\perp) = J_0(k_\perp L)  \;.
\end{equation}
In both the cases the filtering in the image plane is, in a sense, cuts of the power spectrum at the wave number $\sim L^{-1}$, i.e.
\begin{equation}
  \label{eq:approx_convoluton_1_app}
  (2\pi)^4 \int d^2k_\perp\,|\tilde{B}_L(k_\perp)|^2 \sim
  2\pi \int_0^{L^{-1}} dk_\perp\,k_\perp  \;.
\end{equation}

Similarly, the integration over the $k_\parallel$ can be thought as an equivalent to the power spectrum cut-off at the wave number $\sim h^{-1}$:
\begin{equation}
  \label{eq:approx_convoluton_2_app}
  (2\pi)^2 \int dk_\parallel \left| \frac{\tilde{n}(k_\parallel)}{N} \right|^2 \sim \int_0^{h^{-1}} dk_\parallel  \;.
\end{equation}
In fact this assumption means that the characteristic width of the disk must be a well defined quantity, i.e. the disk matter must be well localized in $z$-direction.

We are particularly interested in cases when the filter scale $L$ is less than the largest turbulent scale $h_\mathrm{turb}$. Then, taking account of (\ref{eq:turbulent_velocity_variance_app}), (\ref{eq:approx_convoluton_1_app}) and (\ref{eq:approx_convoluton_2_app}), we can rewrite (\ref{eq:fluctuations_level_app}) as
\begin{align}
  U_L
  &{}= C \sigma_\mathrm{turb}^2
    + (2\pi)^6 \int_{k > h_\mathrm{turb}^{-1}}^{k < L^{-1}}
      d^3k\,|\tilde{G}_L(\vecb{k})|^2 P(k)  \\
  &{}= C \sigma_\mathrm{turb}^2
    + \frac{2\pi}{h} \int_{h_\mathrm{turb}^{-1}}^{L^{-1}} dk\,k P(k)  \\
  &{}= \sigma_\mathrm{turb}^2
    \,\frac{\alpha-1}{2\alpha}\,\frac{h_\mathrm{turb}}{h}\,(1-C)  \\
    &\qquad \times \left[ \frac{2\alpha}{\alpha-1}\,\frac{h}{h_\mathrm{turb}}\,\frac{C}{1-C}
      + 1 - \left( \frac{L}{h_\mathrm{turb}} \right)^\alpha \right]  \;.
  \label{eq:fluctuations_level_approx_app}
\end{align}
As one can see, if we increase the filter width (i.e. directivity pattern), the intensity fluctuations in the line profile decrease. If the filter scale exceeds the largest turbulent scale the level of fluctuations takes its minimum value $U_L \equiv C \sigma_\mathrm{turb}^2$ that does not any more depend on the filter width.


\bsp  
\label{lastpage}
\end{document}